\documentclass[aps,pra,twocolumn,floatfix,showpacs,8pt]{revtex4-1}
\usepackage[utf8x]{inputenc}
\usepackage{textcomp}
\usepackage{amsmath,amssymb,latexsym,graphicx}
\usepackage{mathtools}
\usepackage{overpic}
\usepackage[FIGTOPCAP,scriptsize]{subfigure}
\usepackage{natbib}
\usepackage{url}
\usepackage{hyperref}
\usepackage[all]{hypcap}
\usepackage{multirow}
\usepackage{bm}
\usepackage{relsize}
\usepackage{mathrsfs}
\newcommand*{\diff}{\mathop{}\!\mathrm{d}}

\begin{document}
\title{Trapped Ultracold Bosons in Periodically Modulated Lattices}
\author{Jia-Wei Huo}
\author{Fu-Chun Zhang}
\affiliation{Department of Physics, The University of Hong Kong, Pokfulam Road, Hong Kong}
\author{Weiqiang Chen}
\email{wqchen@hku.hk}
\affiliation{Department of Physics, The University of Hong Kong, Pokfulam Road, Hong Kong\\
and Department of Physics, South University of Science and Technology of China, Shenzhen, China}
\author{M. Troyer}
\affiliation{Institute for Theoretical Physics, ETH Zurich, CH-8093 Zurich, Switzerland}
\author{U. Schollw\"ock}
\affiliation{Physics Department and Arnold Sommerfeld Center for Theoretical Physics, Ludwig-Maximilians-Universit\"at M\"unchen, D-80333 M\"unchen, Germany} 
\date{\today}
\begin{abstract}
Motivated by the recent rapid development of the field of quantum gases in optical lattices, we present a comprehensive study of the spectrum of ultracold atoms in a one-dimensional optical lattice subjected to a periodic lattice modulation. Using the time-dependent density matrix renormalization group method, we study the dynamical response due to lattice modulations in different quantum phases of the system with varying density. For the Mott insulating state, we identify several excitation processes, which provide important information about the density profile of the gases. For the superfluid, the dynamical response can be well described in a local density approximation. This simplification can be valuable in understanding the strong-correlated superfluid in a slow-varying harmonic potential. All these spectroscopic features of an inhomogeneous system can be used as a test for the validity of the Bose-Hubbard model in a parabolic trapping potential.
\end{abstract}
\maketitle
\section{Introduction}
Following the rapid development of experimental techniques for the manipulation and detection of dilute ultracold atom gases, a wide range of fundamental quantum many-body phenomena have been observed. Specifically, due to techniques including Feshbach resonances\cite{Inouye1998} and optical lattices\cite{Orzel2001}, bosonic systems loaded into a periodic lattice described by the Bose-Hubbard model\cite{Fisher1989,Jaksch1998,Giamarchi1987,Giamarchi1988} have been experimentally accessible both in the weakly and strongly interacting regimes with highly controllable parameters, allowing for example the observation of the superfluid to Mott-insulator phase transition driven by quantum fluctuations\cite{Greiner2002}. This achievement has provided a new platform to study quantum many-body physics by virtue of the high degree of control and tuning available\cite{Bloch2008}.

Particularly rich quantum physics is to be expected in the context of quantum many-body physics far from equilibrium, but this remains largely unexplored at the moment. One main experimental difficulty lies in the limited number of techniques of measurement in the strongly correlated system. In this paper, we will focus on one particular technique and point the way to the extraction of additional theoretical information from the raw data, the periodic lattice modulation approach by  St\"{o}ferle {\it et al.}\cite{Stoferle2004} They developed this technique to study the excitation spectrum of the bosonic system in an optical lattice. It acts at a probe with a specific frequency on the ultracold bosons and can be used to reveal the excitation spectrum of the system. More recently, this technique has been widely used in the dynamical control\cite{Chen2011} and the realization of the quantum phase transition\cite{Eckardt2005,Zenesini2009,Struck2011} in optical lattices.

Previous theoretical and numerical studies on this experiment have shown how to extract important information about the system. Theoretically, the technique has been studied via perturbative methods in two limits, by a linear response analysis\cite{Reischl2005,Iucci2006} in the Mott insulating phase and by solving the Gross-Pitaevskii equation in the superfluid regime\cite{Kramer2005}. A drawback of these perturbative methods is that they cannot be used to deal with the whole interaction regime. Numerically, the time-dependent density-matrix renormalization group technique (t-DMRG) has been applied to simulate the experimental setup in a quasi-exact fashion\cite{Kollath2006,Clark2006,Poletti2011}. Basic features seen in the experiment could be reproduced successfully. 

Although these studies have opened a window on the understanding of the experimental observation\cite{Stoferle2004}, further theoretical and numerical questions about the excitation spectroscopy arise mainly for three reasons. First, ultracold experiments are carried out in a harmonic trapping potential. Although it had been pointed out that this will induce spectral broadening and a shifting of peaks, a detailed and quantitative study of this issue is still lacking. This is important to assess whether the presence of the trap qualitatively or quantitatively changes the behavior of the homogeneous system. Secondly, a direct comparison to the experimental observation, which averages over many 1D systems with different particle numbers and densities per 1D tube, is not satisfactory unless the dependence on the density is taken into account. Therefore, building upon previous studies, this paper will generally focus on extracting new information from the spectrum due to the harmonic confinement and different densities. 

\section{The Bose-Hubbard model}\label{sec:model}
Ultracold bosons in an optical lattice can be described by the Bose-Hubbard model\cite{Fisher1989,Jaksch1998}
\begin{equation}
  \hat{H}=-J\sum_{j}(\hat{b}^\dag_{j}\hat{b}^{}_{j+1}+\text{H.c})+\frac{U}{2}\sum_{j}\hat{n}^{}_{j}(\hat{n}^{}_{j}-1)+\sum_{j}V_j\hat{n}_{j},
\end{equation}
where $\hat{b}_j$ and $\hat{n}_j$ are the annihilation and number operators on site $j$, respectively. The first term describes the hopping process between nearest neighbours, while the second one depicts the on-site interaction. The last term models the harmonic trapping potential $V_j = V_t(j-j_0)^2$ with $V_t$ the curvature and $j_0$ the center of the system.
In this paper, we are only interested in the absorption spectrum of ultracold bosons in a 1D optical lattice. We assume the 1D optical lattices all to be directed along the $x$-direction. We denote by $V_x$ and $V_{\bot}$ the laser strength along the $x$-direction and $yz$-directions respectively. For deep lattices, the hopping matrix element $J$ and on-site interaction $U$ can be approximated as\cite{Zwerger2003,Kollath2006}
\begin{eqnarray}
  \frac{J}{E_r}&=&\frac{4}{\sqrt{\pi}}\left(\frac{V_x}{E_r}\right)^{\frac{3}{4}}\exp\left(-2\sqrt{\frac{V_x}{E_r}}\right)
\end{eqnarray}
and
\begin{eqnarray}
  \frac{U}{E_r}&=&4\sqrt{2\pi}\frac{a_s}{\lambda}\sqrt[4]{\frac{V_xV^2_\bot}{E^3_r}},
\end{eqnarray}
where $E_r$ is the recoil energy, $a_s$ is the s-wave scattering length, and $\lambda$ is the wavelength of the laser forming the optical lattice.  The parameters used in the calculations are $a_s\!=\!5.45$ nm, $\lambda\!=\!825$ nm, $V_{\bot}\!=\!30E_r$.  In order to investigate the absorption spectrum, one applies a sinusoidal modulation of the $x$-direction laser strength $V_x$ starting at $t = 0$ with frequency $\omega$ and amplitude $\delta V$, {\it i.e.}, $V_x(t\!=\!0)=V_0$ and $V_x(t\!>\!0)=V_0+\delta V\sin\omega t$\footnote{The change of the phase boundary due to the modulation discussed in Ref. \cite{Eckardt2005} is negligible in this form of modulation as long as its strength is not strong enough.}, and measures the energy absorbed by the system. The absorbed energy is strongly frequency-dependent and gives information about the excitation spectrum.

\section{Methods}\label{sec:method}
\subsection{Time-dependent perturbation (t-perturbation)}\label{subsec:perturbation}
If the modulation amplitude $\delta V$ is small and the system is in a deep Mott-insulating state ($J\!\ll\!U$), the system can be understood within the framework of time-dependent perturbation theory where the Hamiltonian reads
\begin{eqnarray}
  \hat{H}[J(t),U(t)]\approx\hat{H}_0+\hat{H}'(t),
\end{eqnarray}
where $\hat{H}_0\!\!\equiv\!\!\hat{H}(t\!\!=\!\!0)$. (For brevity $U_0\!\!\equiv\!\!U(t\!\!=\!\!0)$ and $J_0\!\!\equiv\!\!J(t\!\!=\!\!0)$ are used hereafter.) To keep the only term contributing to the excitations, we make a transformation\cite{Reischl2005}
\begin{eqnarray}
  \hat{H}'(t)\rightarrow\widetilde{\hat{H}'}(t)=\hat{H}'(t)-\frac{\delta U}{U_0}H_0.
\end{eqnarray}
By further neglecting the time-independent term, we have
\begin{equation}
  \widetilde{\hat{H}'}(t)\!=\!-F_J\sin\omega t\sum_{j=1}^{L-1}(\hat{b}^\dag_{j}\hat{b}^{}_{j+1}+\text{H.c.}),
\end{equation}
with the coupling constant
\begin{equation}
 F_J=\left.\left(\frac{\partial\ln J}{\partial V_x}-\frac{\partial\ln U}{\partial V_x}\right)\right|_{V_x=V_0}J_0\delta V. 
\end{equation}
This coupling constant has been shown to be valid in the large $U$-limit\cite{Iucci2006}.

In standard time-dependent perturbation theory, the transition probability is given by
\begin{multline}
  W_{mn}(t)=\frac{|H_{mn}'|^2}{4\hbar^2}\left|\frac{1-e^{i(\omega_{mn}+\omega)t}}{\omega_{mn}+\omega}\right.\\
  \left.-\frac{1-e^{i(\omega_{mn}-\omega)t}}{\omega_{mn}-\omega}\right|^2.
\end{multline}
Here $H_{mn}'$ is the matrix element of $\hat{H}'$ between two eigenstates $m$ and $n$ of the unperturbed Hamiltonian $\hat{H}_0$.
So the energy absorbed is
\begin{equation}
  \Delta E(t)=\sum_{m}\hbar\omega_{m0}W_{m0}(t).
\end{equation}

\subsection{t-DMRG}\label{subsec:tdmrg}
Numerically, we use the t-DMRG method to study the time evolution of the system\cite{Daley2004,White2004,Schollwock2011}. This method is a quasi-exact algorithm which allows for simulating real time evolutions of 1D quantum many-body systems, which operates on a class of matrix product states\cite{Vidal2003,Vidal2004,Verstraete2004}. To begin with, a conventional finite-system DMRG algorithm is used to determine the ground state, $|\psi(t=0)\rangle$, of the Hamiltonian at time $t=0$, $\hat{H}(t=0)$ for a system with $L$ sites and $N$ bosons. Then a full time evolution of the quantum state, $|\psi(t)\rangle$, is calculated with the t-DMRG algorithm based on a Trotter decomposition of time steps. We keep up to 200 states in the reduced Hilbert space in the algorithm. In order to reduce the error from the Trotter decomposition, we use a linear fit to extrapolate the results to Trotter time steps $\delta t\rightarrow 0$. Convergence in the number of states kept has also been checked, such that on the time scales simulated the results are quasi-exact.

The full time dependence of the total energy reads
\begin{eqnarray}
  E(t)&=&\langle\psi(t)|\hat{H}(t)|\psi(t)\rangle \nonumber \\
  &\approx&\langle\psi(t)|\hat{H}_0|\psi(t)\rangle+\langle\psi(t)|\hat{H}'(t)|\psi(t)\rangle.
\end{eqnarray}
The main contribution to the energy transfer to the system comes from the first term as the time average of $H'(t)\propto\sin\omega t$ vanishes. Thus, to get the absorption spectrum, we calculate the energy absorbed up to a given time $t_m$,
\begin{equation}
  \Delta E=\langle\psi(t=t_m)|\hat{H}(t=0)|\psi(t=t_m)\rangle-E_0,
\end{equation}
where $E_0$ is the ground state energy\cite{Clark2006}. This method is essentially equivalent to fitting the full time dependence of the energy curve\cite{Kollath2006}.

\section{Results}\label{sec:results}
\subsection{Absorption spectroscopy with Mott domains}\label{subsec:mott}

The Mott phase in the Bose-Hubbard model is characterized by short-ranged exponentially decaying correlations and commensurate filling. 
In a trapped system, there is no homogeneous Mott phase because of the harmonic trap. But for large enough $U$ and suitable particle filling, the system may still have one or several Mott domains separated by superfluid domains in space\cite{Kollath2006,Clark2006}. The number of Mott domains and the filling in those domains depend on the average density of the system. In the following, we will study the absorption spectroscopy in the presence of Mott domains with varying densities. 
\begin{figure}[ht]
  \begin{center}
    \subfigure[]{
      \includegraphics[width=0.45\textwidth]{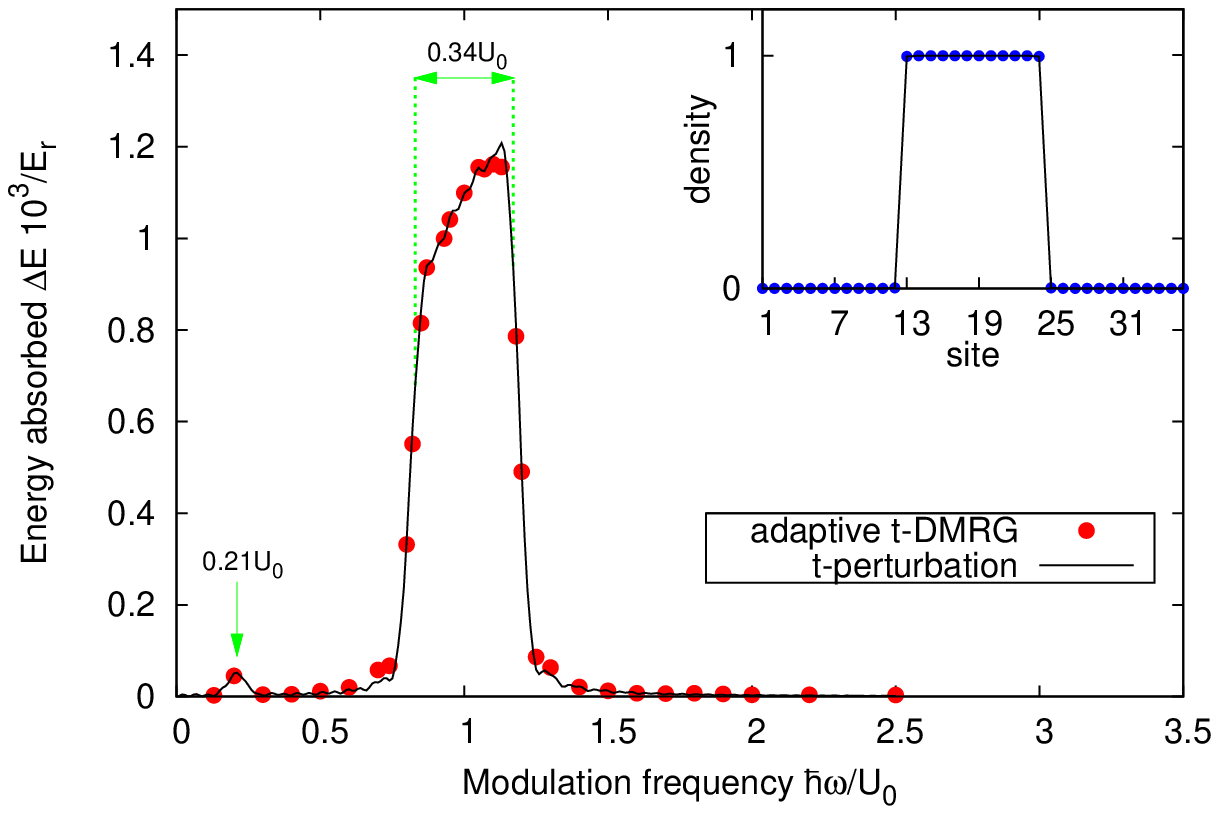}
    }
    \subfigure[]{
      \includegraphics[width=0.20\textwidth]{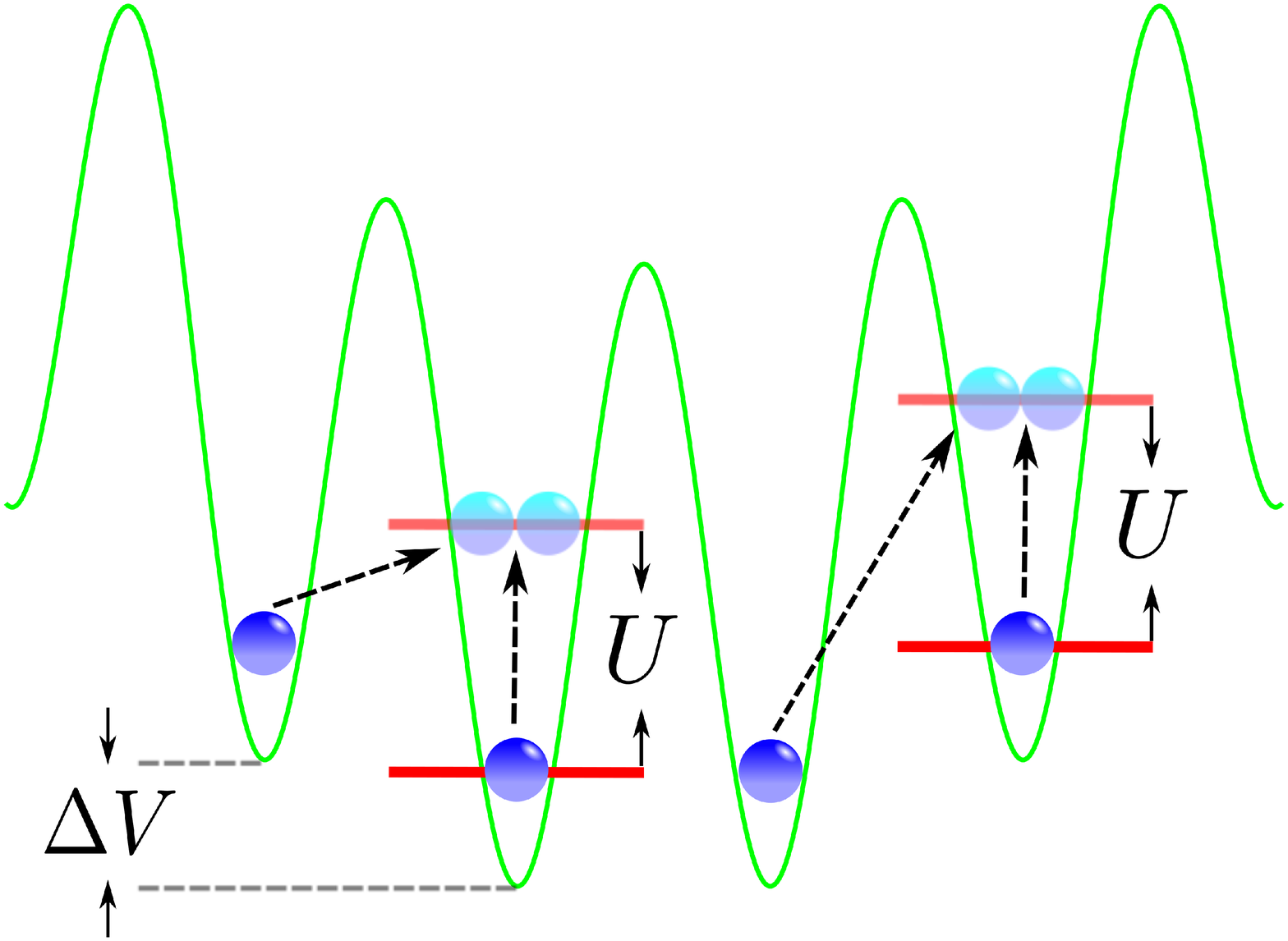}
    }
    \subfigure[]{
      \includegraphics[width=0.20\textwidth]{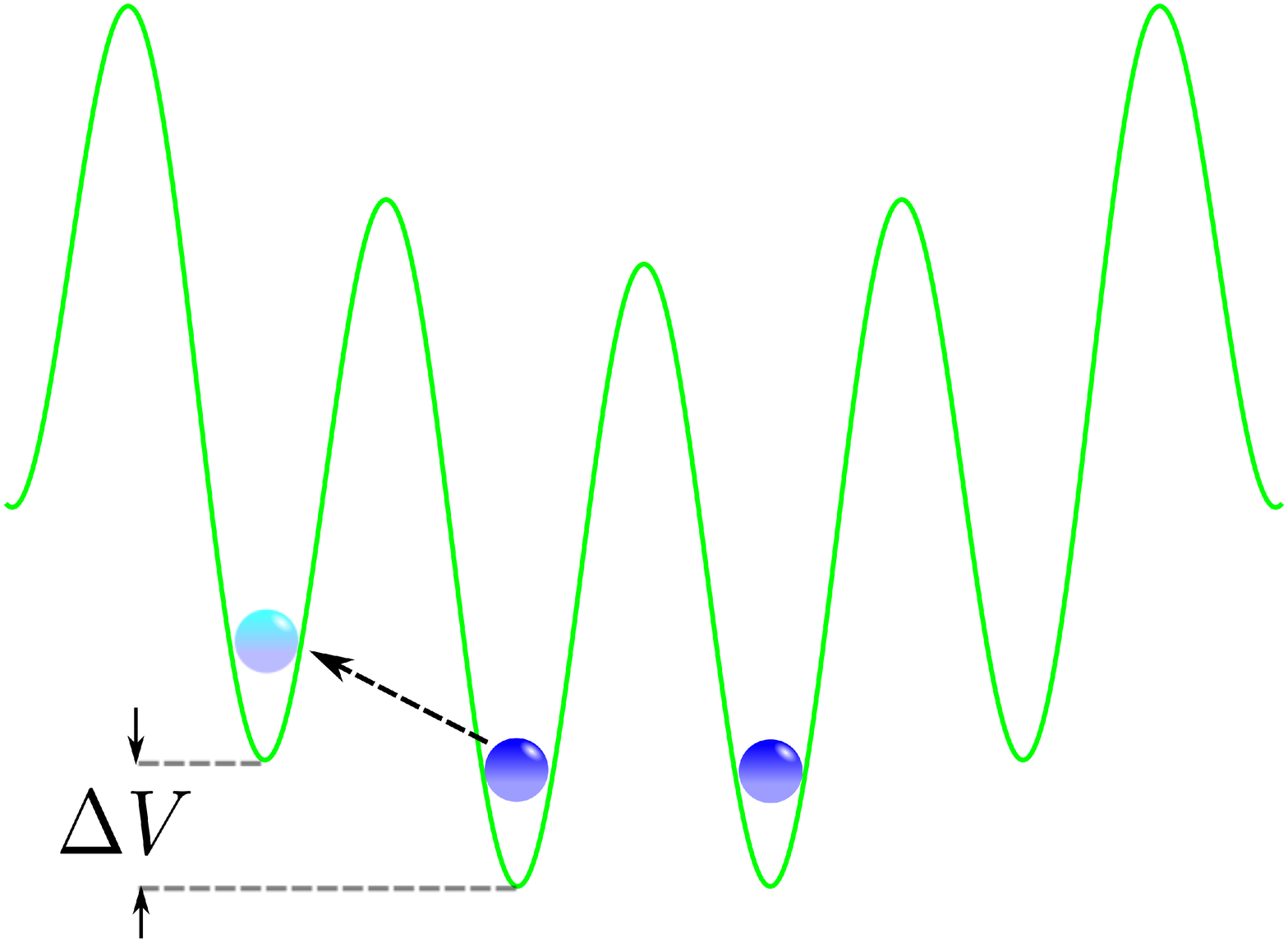}
    }
    \caption{\label{mott1}(Color online) (a) Total energy absorbed as a function of the modulation frequency $\omega$ subjected to a small modulation $\delta V=0.01V_0$. Both the results of t-DMRG and time-dependent perturbation theory are shown. The inset shows the initial density distribution of the system. (b) and (c) are schematic descriptions of the excitations due to the modulation.}
  \end{center}
\end{figure}

For the low density case, we consider a system with 12 bosons in a deep optical lattice with $V_0=15E_r$ and 36 sites. The curvature of the trapping potential is $V_t=0.0123 E_r\approx 0.017 U_0$. The density profile of the system is shown in the inset of Fig.~\hyperref[mott1]{1(a)}, where there is only one Mott domain with unit filling in the center of the system. Then we consider the absorption spectrum which is measured at time $t_m = 100\hbar/E_r$ and depicted in Fig.~\hyperref[mott1]{1(a)}. In a homogeneous Mott phase, the absorption spectrum is highlighted by a sharp peak at energy $U$\cite{Kollath2006}. This resonance, corresponding to a particle-hole excitation, is sharp since the excited states are almost degenerate. However, the situation is different after a trapping potential is applied. The trap introduces a difference in potential energy between two neighboring sites
\begin{equation}\label{eqn:diff}
  \Delta V(j,j\pm 1)=V_{t}[1 \pm 2(j-j_0)].
\end{equation}
As a rough estimate, the particle-hole excitation energy from site $j$ to $j+1$, as shown in Fig.~\hyperref[mott1]{(1b)},  will deviate from $U$ by $V(j,j+1)$. The width of the peak is expected to be determined by the potential difference at the edges of the domain where the potential energy is maximized. According to the density profile shown in the inset of Fig.~\hyperref[mott1]{1(a)}, the edge is at site 13 and site 24 where the potential difference is $0.17U_0$. So the estimated width of the $1U$ peak is $0.34U_0$, which coincides very well with our numerical results in Fig.~\hyperref[mott1]{1(a)}. 

Another new feature in the spectrum is the small peak at $\hbar\omega \approx 0.21U_0$ which corresponds to the particle-hole excitation where a particle hops to an empty site as shown in Fig.~\hyperref[mott1]{1(c)}. This stems mainly from hopping from site 13 to site 12 and site 24 to site 25, where the potential difference is $0.204 U_0$, in excellent agreement with numerics.

\begin{figure}[ht]
  \centering
  \subfigure[]{
    \includegraphics[width=0.45\textwidth]{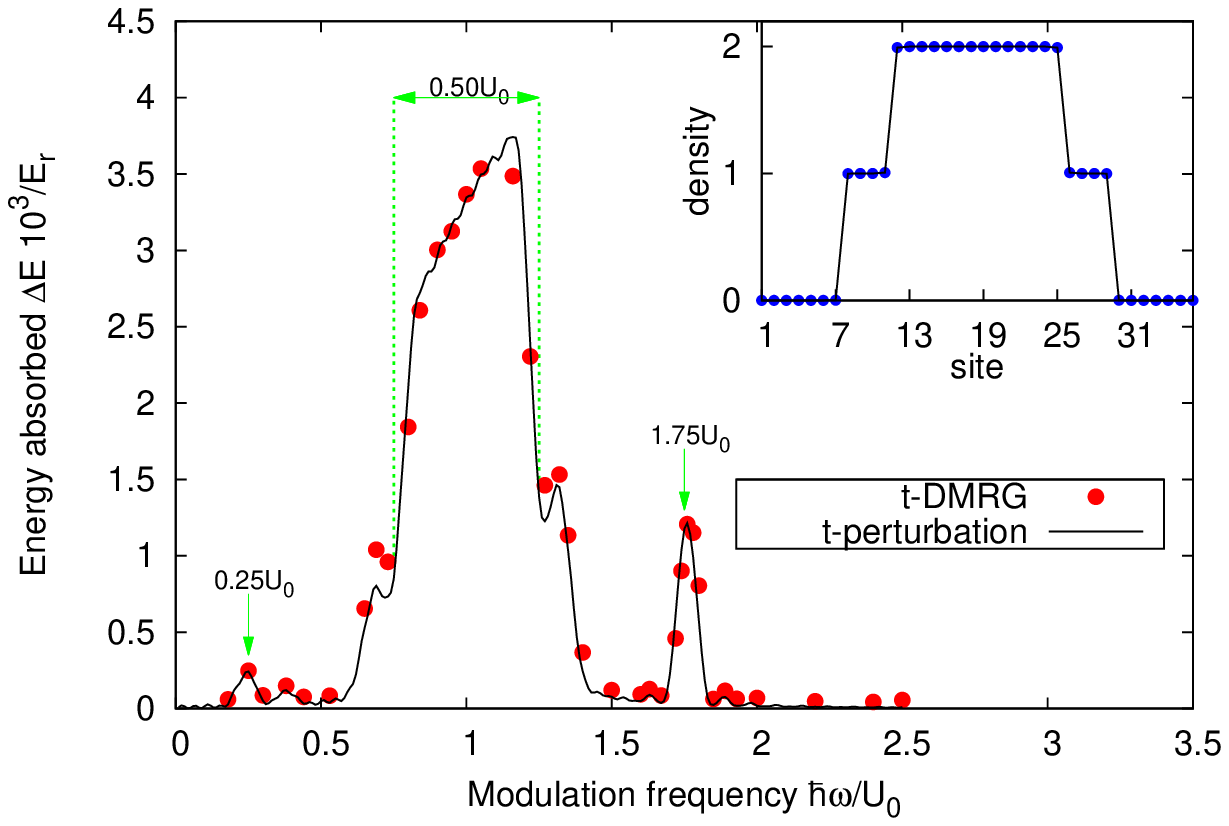}
  }
  \subfigure[]{
    \includegraphics[width=0.20\textwidth]{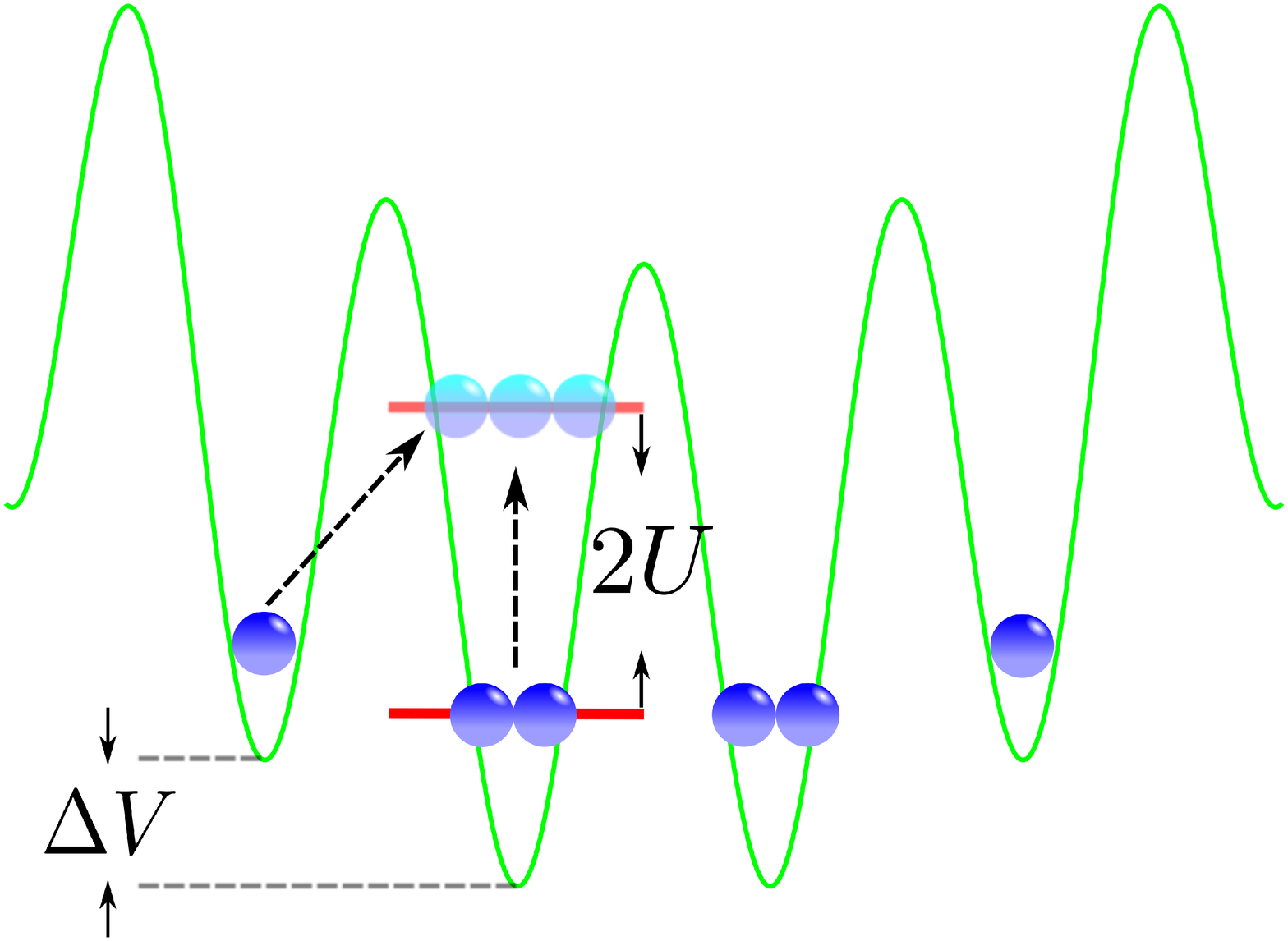}
  }
  \subfigure[]{
    \includegraphics[width=0.20\textwidth]{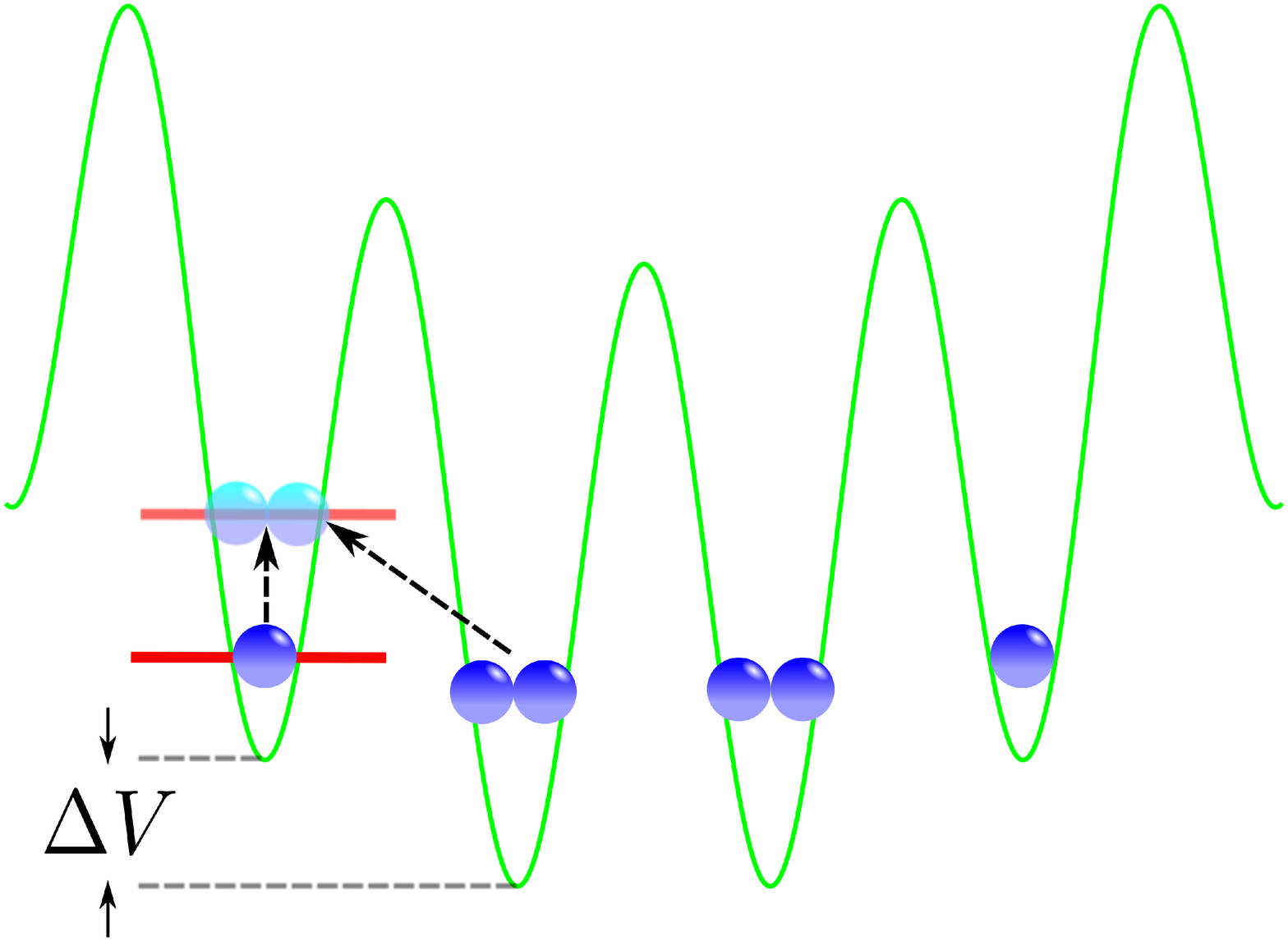}
  }
  \caption{\label{mott2}(Color online) Absorption spectrum (a) in a confined system with 36 particles. The inset of (a) shows the initial density profile. (b) and (c) are schematic descriptions of the excitations between two different Mott domains. (b) describes the second predominant absorption peak at $\hbar\omega=1.75U_0$ in (a), while (c) is used to explain the small peak at $\hbar\omega=0.25U_0$. From the inset one can calculate the chemical potential difference at the Mott domain walls via Eq.~\ref{eqn:diff}, which gives $\Delta V=\Delta V(j,j+1)=\Delta V(12,11)=\Delta V(25,26)=0.25U_0$.}
\end{figure}

As the particle number $N$ increases, there will be several Mott domains with different fillings as shown in the inset of Fig.~\hyperref[mott2]{2(a)} where the particle number is 36 and the other parameters are the same as in the previous case. Besides the broadened $1U$ peak, there are  two more peaks: one centered at 0.25U$_0$, and another centered at $1.75U_0$. The former corresponds to the hopping of electrons between two Mott domains with different doping at the domain boundaries as depicted in Fig.~\hyperref[mott2]{2}, and its frequency is determined by the potential energy difference $\Delta V(12,11)=\Delta V(25,26)=0.25U_0$. 

At a first glance, the peak at $1.75 U_0$ originates from the hopping of bosons from the unit filled Mott domain into the doubly occupied region. This is qualitatively but not quantitatively accurate.  If there is no external potential, the excitation energy is exactly $2U$, indicating the non-unit filling of the system\cite{Kollath2006}. However, due to the parabolic potential, one has to take into account the difference of the potential energy of the two site involved in the hopping process $\Delta V$ as shown in Fig.~\hyperref[mott2]{2(a)}. Thus, the position of the excitation peak should be $\hbar\omega=2U-\Delta V$. Both excitations in Fig.~\hyperref[mott2]{2(b)} and Fig.~\hyperref[mott2]{2(c)} involve exactly the same two sites in the system, so we have $\Delta V = 0.25 U_0$ which is the peak energy analyzed above. Then the correct position should be at $1.75 U_0$ which is exactly the results in our numerical calculations. This shift has also been observed by St\"{o}ferle \textit{et al.}, who reported that there was a peak at about $1.9U$ in the Mott phase\cite{Stoferle2004}.

An interesting consequence of these observations is that by combining the broadening effect of the $1U$ peak and the shift of the $2U$ peak, one can directly determine $U$. In the case considered here, we find
\begin{equation} \label{eqn:appx}
  U\approx\frac{1}{2}(\frac{W_1}{2}+U_2),
\end{equation}
with $W_1$ the width of the $1U$ peak and $U_2$ the position of the $2U$ peak. This formula works because the broadening and shifting effect are caused by almost the same chemical potential difference $\Delta V$. This is useful in the calibration of the on-site interaction parameter $U$.

\begin{figure}[ht]
  \begin{center}
    \subfigure[]{
      \includegraphics[width=0.45\textwidth]{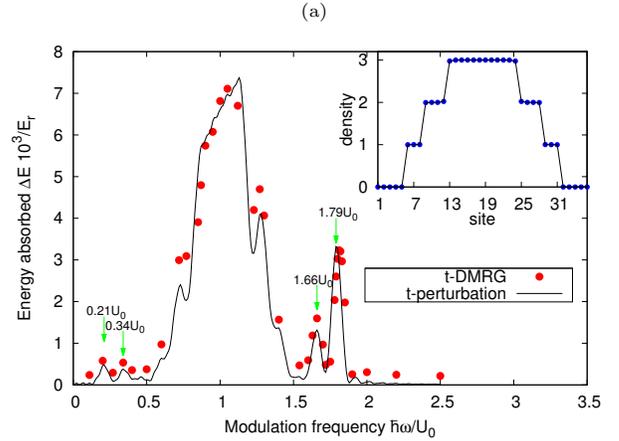}
    }
    \subfigure[]{
      \includegraphics[width=0.45\textwidth]{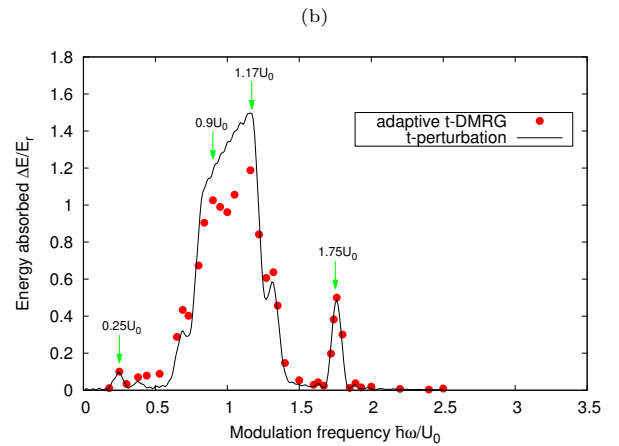}
    }
    \caption{\label{mott3}(Color online) For a weak modulation of a large system with 58 particles we have (a) the total energy absorbed vs. the modulation frequency. On the other hand, in (b) we show the spectroscopy of the system in Fig.\hyperref[mott2]{2} under a strong modulation $\delta V=0.2V_0$.}
  \end{center}
\end{figure}

In order to further corroborate the connection between the energy shift and the density profile, we performed an additional calculation in a Mott system involving triple occupancies shown in Fig.~\hyperref[mott3]{3(a)}. In this high-density system, there exist particle-hole excitations at the various Mott domain walls. Since the chemical potential difference is spatially varying, the shifts away from $2U$ are different. Therefore, we can see the $\Delta V$ peak and $2U$ peak both split into two peaks due to this difference.

To compare our results with  experiments where the modulation amplitude is up to $20\%$ of the lattice depth, we also carried out a simulation for a large modulation (Fig.~\hyperref[mott3]{3(b)}). Here the breakdown of perturbation theory indicates the saturation effect in a real system. Due to this effect and the relatively large trap curvature the splitting of the $1U$ peak is pronounced.  We can still identify the positions of the excitations from the shifts. For example, the peak at $1.17U_0$ is related to the hopping process from site 14 to 13, or from site 23 to 24. Another important finding is that the positions of the $\Delta V$ peak and $2U$ peak are robust. Also, the saturation effect for these excitations is less significant than that of the $1U$ resonance.

Based on the findings above, the position of the $2U$ peak reveals important information of the density profile of the Mott system. For one thing, the number of the peaks indicates the number of Mott layers in the ``wedding cake'' structure. For another, the shift away from $2U$ provides important information about the positions of the Mott domain walls.

\subsection{Absorption spectrum in a superfluid}\label{subsec:superfluid}

In this section, we turn to the superfluid state, where the interpretation of the excitation spectroscopy is less straightforward than in the Mott regime. Without loss of generality, we choose $V_0=4E_r$ and $\delta V=0.2V_0$ in all the calculations in this subsection, leading to $U/J\approx 5$. In contrast with the Mott insulating state where the hopping is substantially suppressed, both the parameters $J$ and $U$ play an important role in determining the main properties of the superfluid. What makes the situation more complicated is the external harmonic trap, which introduces inhomogeneity into the system. A simplification occurs nevertheless as we will show that one can map the absorption spectrum in a trapped system to the homogeneous one by using the local density approximation (LDA). Mathematically, this means
\begin{equation}\label{eqn:lda}
  \Delta E^{\text{trapped}}(\omega)=\int\rho(\vec{r})\Delta\epsilon^{\text{homo}}(\omega,\rho(\vec{r})) \diff \vec{r},
\end{equation}
where $\rho(\vec{r})$ is the spatially dependent density, $\Delta E^{\text{trapped}}(\omega)$ is the energy transferred as a function of frequency $\omega$ in a trapped system, and $\Delta\epsilon^{\text{homo}}(\omega,\rho(\vec{r}))$ is the energy absorbed density in a homogeneous system with particle density $\rho(\vec{r})$.

\begin{figure}[ht]
  \begin{center}
    \subfigure[]{
      \includegraphics[width=0.45\textwidth]{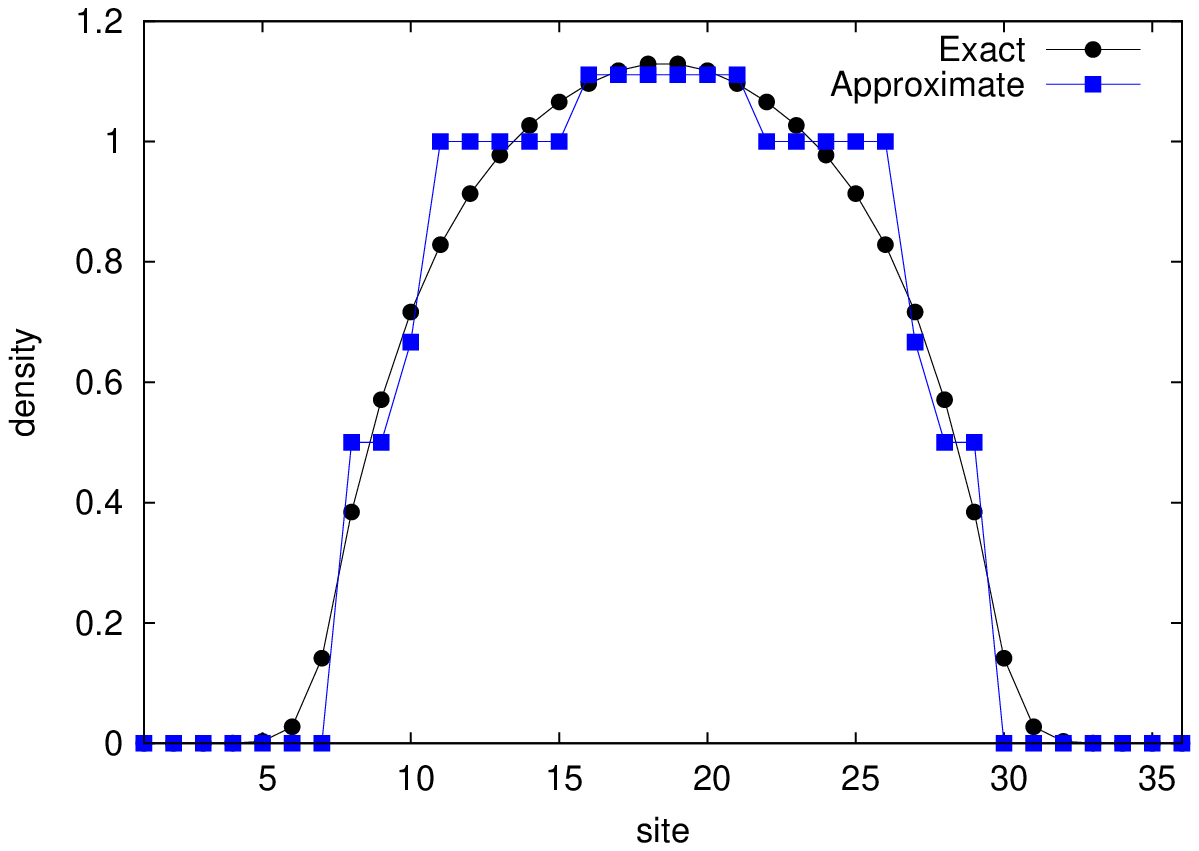}
    }
    \subfigure[]{
      \includegraphics[width=0.45\textwidth]{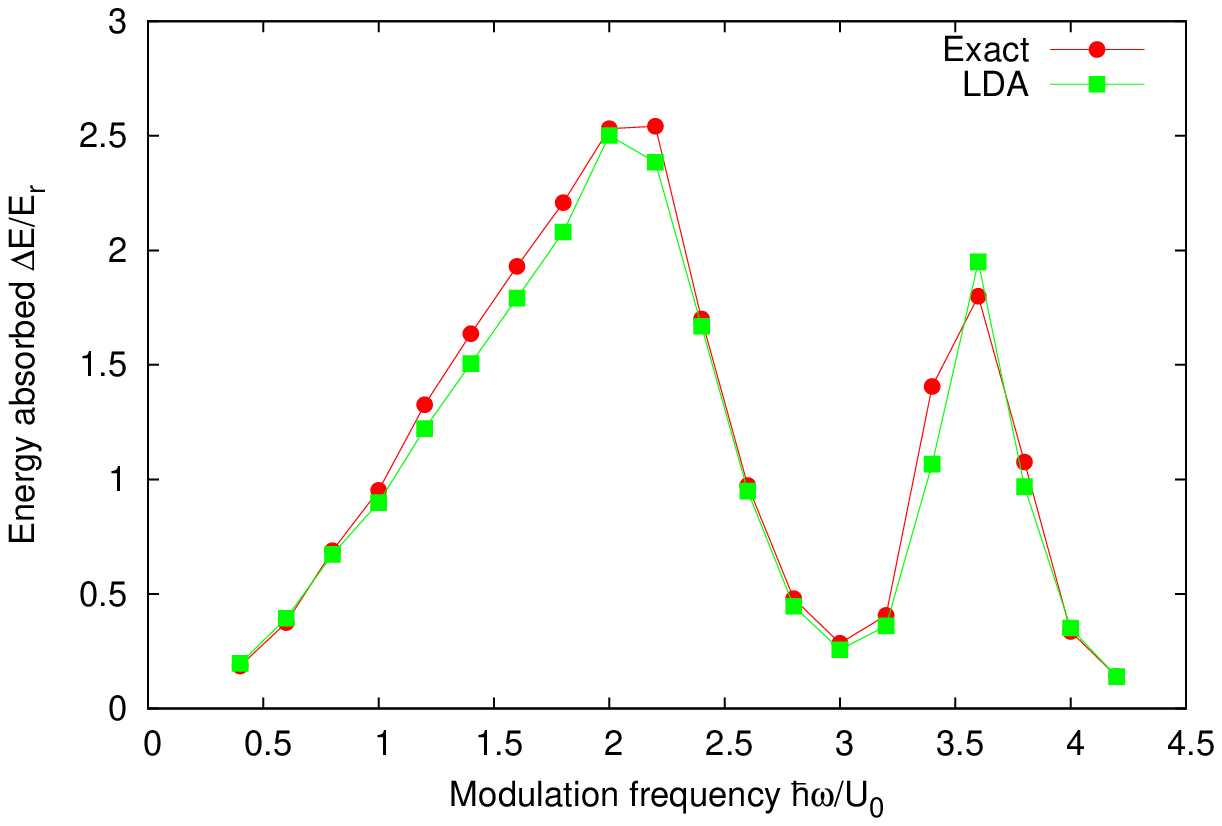}
    }
    \caption{\label{sf1}(Color online) (a) Density profiles for the system. Both the exact (circles) and the approximate (squares) results are shown. (b) The absorption spectroscopy in the form of the exact t-DMRG results (circles) and LDA (squares). The lines are drawn to guide the eye.} 
  \end{center}
\end{figure}

To show this approximation really holds, we compare the exact spectrum of a trapped system with a result from the LDA (Fig.~\hyperref[sf1]{4(b)})\cite{footnote1}. To simplify the calculation of spectra we first make an approximation on the density profile (Fig.~\hyperref[sf1]{4(a)}). Although this approximation seems rough, the resulting spectrum is in good agreement with the exact one. Then the approximate results can be calculated as $\Delta E^{\text{LDA}}\!=\!\sum_i\rho(i)\epsilon^{\text{homo}}(\omega,\rho(i))$, where $\epsilon^{\text{homo}}(\omega,\rho(i))$ is calculated in a homogeneous system with density $\rho(i)$. 

From the comparison, it is clear that the LDA works extremely well despite the approximate density profile. From a physical point of view, the validity of LDA stems from the slow-varying density profile in the harmonically confined superfluid. Also the chosen interaction is away from the location of the phase transition. Thus, here the main effect of the parabolic trap is no more than introducing a slow-varying inhomogeneity. On the other hand, LDA must fail in the Mott insulating phase since there exists non-trivial excitations at the boundaries between Mott domains, as we saw in the previous section.

\begin{figure}[ht]
  \centering
  \subfigure[Homogeneous]{
    \includegraphics[width=0.45\textwidth]{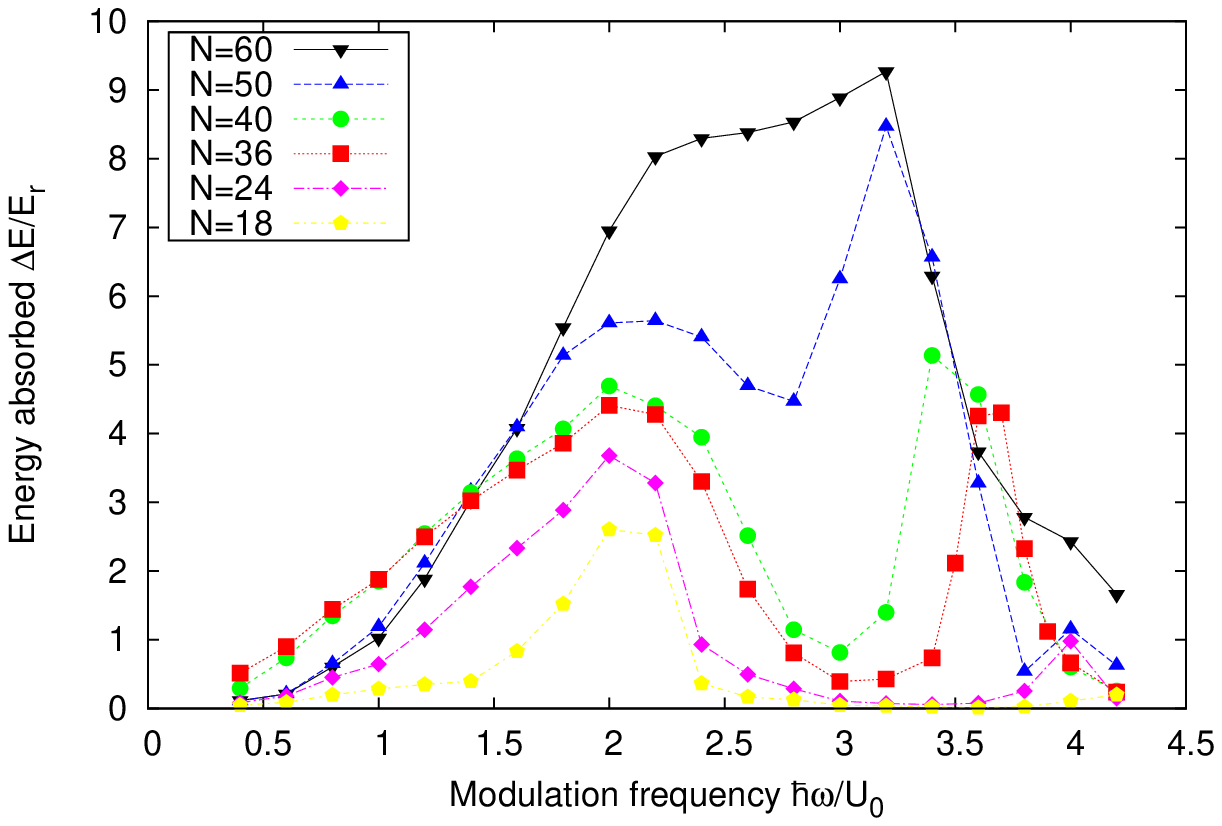}
  }
  \subfigure[Confined]{
    \includegraphics[width=0.45\textwidth]{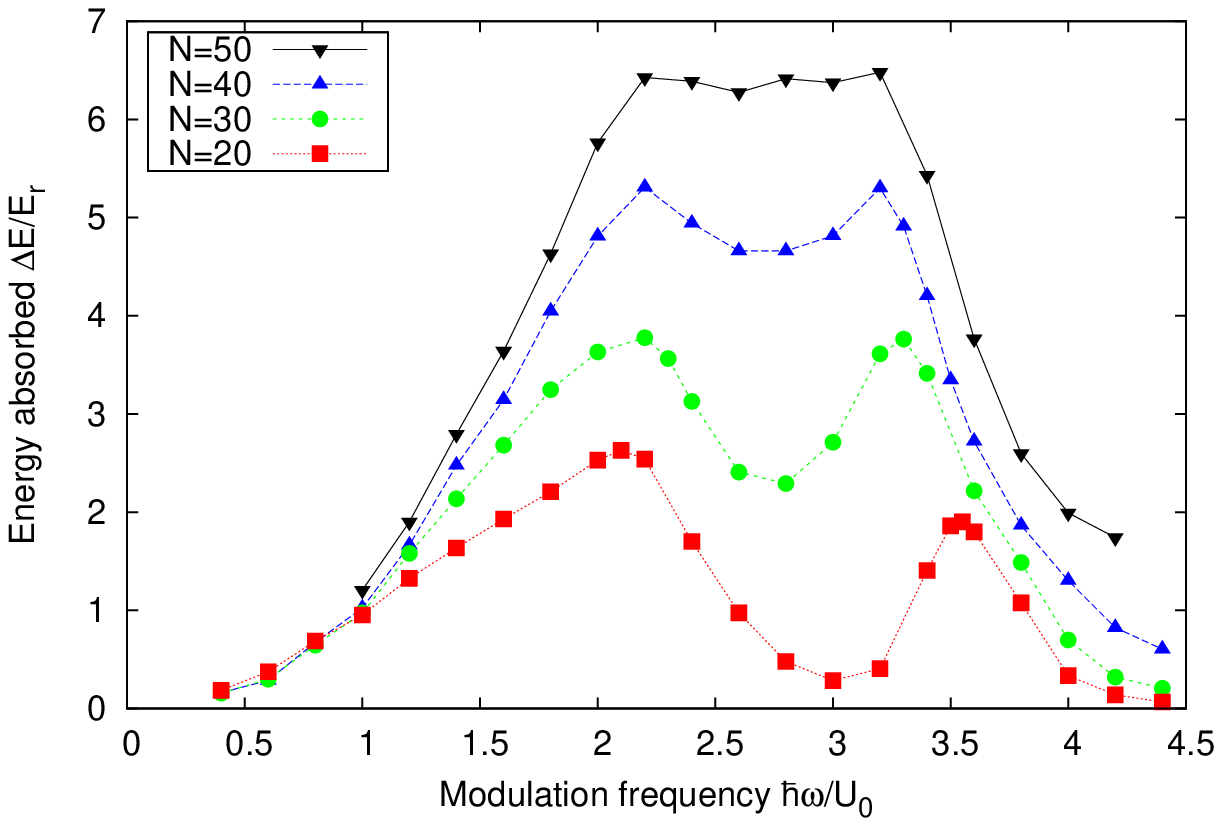}
  }
  \caption{\label{sf2}(Color online) Spectra with different particle number $N$ in (a) a homogeneous system and (b) trapped one. The system size $L=36$ is fixed.}
\end{figure}

The significance of the LDA is that one can understand the properties of a trapped superfluid with the help of a homogeneous one. To further test the LDA and to study how the density influences the spectroscopy, we performed calculations for different particle numbers both in homogeneous and confined systems (see Fig.~\hyperref[sf2]{5}). 

For the unconfined atoms, there is a sharp excitation between $3U$ and $4U$ for intermediate densities. This resonance can be interpreted as two particle-hole excitations with both particles on the same site\cite{Clark2006}. Here, we find that this peak is very sensitive to the density of the quantum gas. As shown in Fig.~\hyperref[sf2]{5(a)}, when the density of the system is below 1/2, this excitation is negligible because it is rare for three particles to be nearest neighbours. As the density increases to unity, {\it i.e.}, $N=36$, the spectrum shows a sharp peak around $3.8U$. Strictly speaking, its excitation energy involves two parts. The first part is the excitation energy to the $3U$-Hubbard band, while the second one comes from the change of the kinetic energy for these two atoms from the delocalized ground state to the localized excited state. In a system with unit filling, the energy difference per atom between delocalization and localization is about $2J$. Therefore, the total energy gained for this type of excitations should be $3U+2\times2J$. In our case where $J\approx 0.2U$, the estimate $3.8U$ coincides with the t-DMRG result. 

When we consider the system in a harmonic trap (see Fig.~\hyperref[sf2]{5(b)}), we find a similar density dependence for the spectrum: in the low-density regime, it exhibits a two-peak structure; as the particle number increases, the two peaks merge and become a broad one. This subtle change is also an indication for the density of the quantum gases. 

\begin{figure}[ht]
  \centering
  \subfigure[Homogeneous]{
    \includegraphics[width=0.45\textwidth]{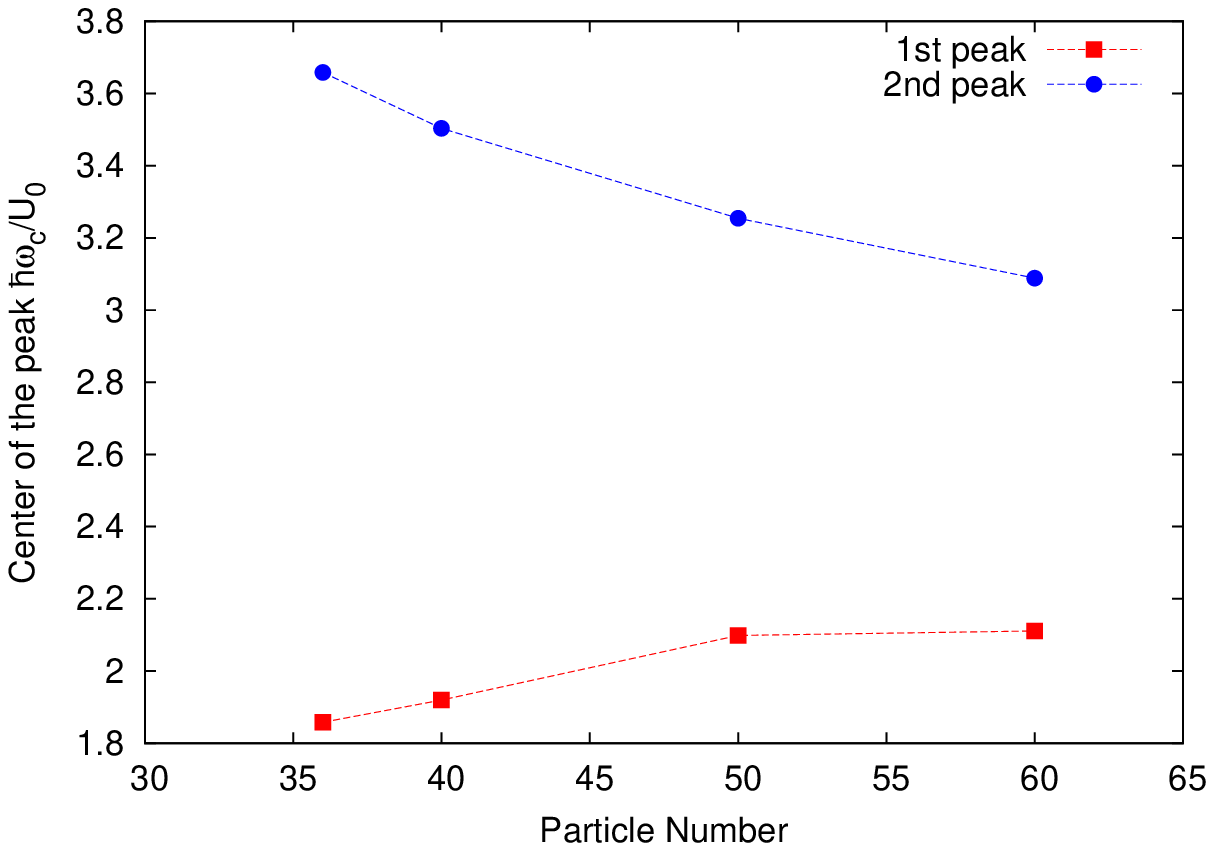}
  }
  \subfigure[Confined]{
    \includegraphics[width=0.45\textwidth]{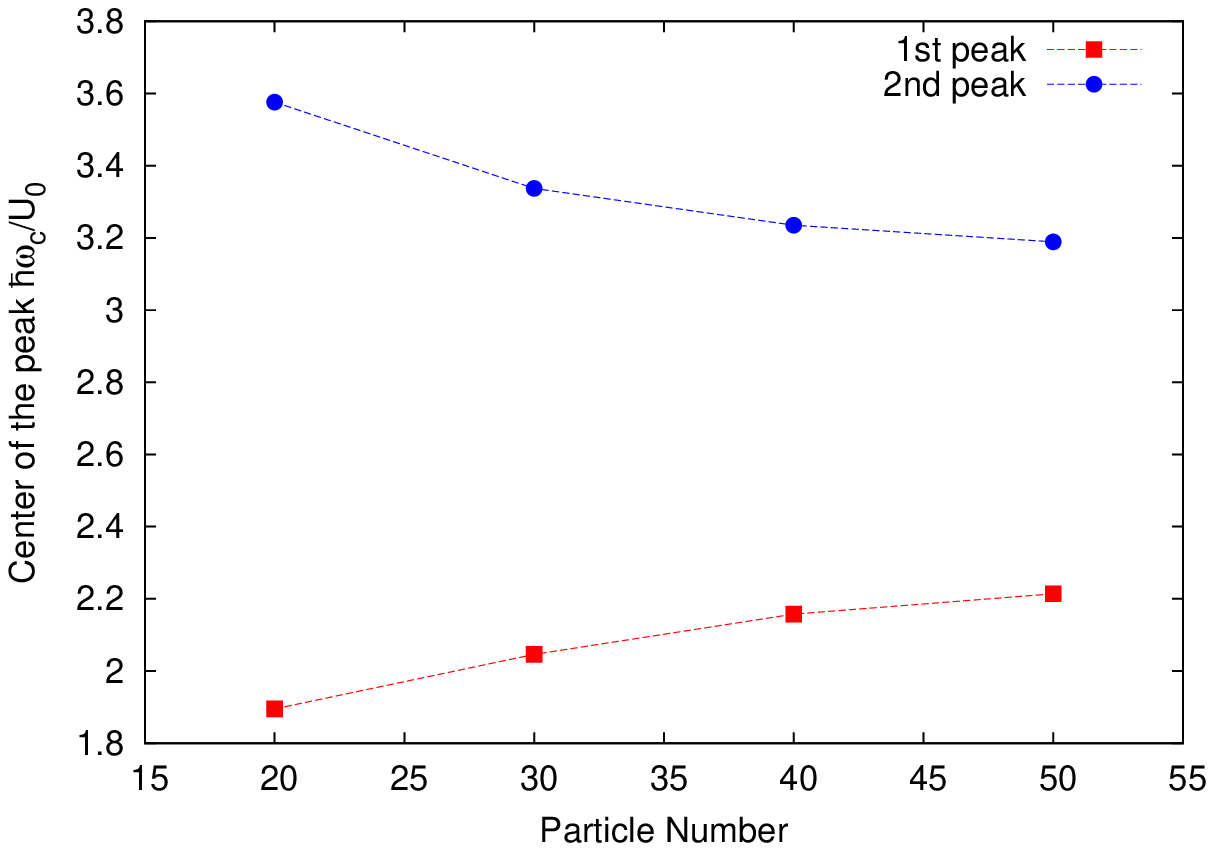}
  }
  \caption{\label{sf3}(Color online) The centers of the peaks in the spectra of Fig.~\hyperref[sf2]{5} for (a) a homogeneous system and (b) trapped one. The values are obtained by fitting the curves with bimodal gaussian distributions.}
\end{figure}

To demonstrate the resemblance of the two cases (absence or presence of a trap), we also plot the centers of two peaks in the spectra as a function of the number of particles in Fig.~\hyperref[sf3]{6}. It is clear that the two peak centers shift towards each other as the particle number increases in both cases. For the first peak, it moves towards higher energy because hoppings between sites with differing fillings become important with more particles. For the second one, its excitation energy of the second peak involves $3U$ for the Hubbard-band and $4J$ for the kinetic energy. It shifts towards lower energy because the ground state for atoms is not completed delocalized for a high density, which compromises the energy gained from localization in the excited state. From this argument, it is expected that the peak would be located around $3U$, which is consistent with our numerical result. With further increased density, the $3U$ resonance and the excitations around $2U$ merge together and form a broad continuum eventually.

Thus, the density dependence of the absorption spectrum in either a homogeneous or trapped system is basically the same. To understand this in the framework of the LDA, a trapped system can be divided into several homogeneous region with slow-varying density. The energy transferred $\Delta E^{\text{trapped}}$ is mainly determined by the high-density region since it maintains the most important weight in the integral of Eq.~(\ref{eqn:lda}). Then it is reasonable to map the trapped system to a homogeneous one.

\section{Discussion and Conclusion}\label{sec:discussion}

How can these observations on the absorption spectroscopy in a trapped system be turned into tools for experimental analysis? It turns out that they lead to schemes to test the validity of the Bose-Hubbard Hamiltonian and to calibrate the interaction parameters.

First of all, our findings in the Mott regime are useful in determining the density profile of the cold gases in an optical lattice. As has been mentioned in Sec. \ref{subsec:mott}, the broadening of the 1$U$ excitation and the shift of the 2$U$ resonance are both directly related to the density distribution of the system. For example, the number of the 2$U$ peaks indicates the number of the Mott domains, while their shifts away from 2$U$ can help to calculate the boundaries of the domains. Also, it is straightforward to test the Bose-Hubbard model by comparing the results from the spectroscopy with those from the time-of-flight imaging technique. Another important application in the Mott phase is calibrating the parameter $U$, which has proven to be difficult in a deep lattice\cite{Buchler2010}. The simplest way to do this with the spectroscopy is to fit the first strong $1U$ peak. To improve the accuracy, one can take into account the $2U$ resonance to recalibrate. For instance, the $U$ is directly related to the width of the first peak and the position of the second peak according to Eq.~(\ref{eqn:appx}). The advantage of this method is that no other fit parameters are needed. 

For the superfluid case, the main concern is how to make use of the LDA. Since one can map the absorption spectrum of a confined system to a homogeneous one, it suggests that the basic features of the superfluid, including the ground state and the excited states, can be well described by the LDA. This is an important characteristic to distinguish different phases driven by quantum fluctuations. Whereas we only test this approximation in 1D, it is quite reasonable to conclude that it works also in 2D and 3D where fluctuations are less important. This generalization would greatly simplify many theoretical studies on the actual superfluid since it builds a bridge connecting a confined system and a homogeneous one. This  can also be used as a criterion to test the validity of Bose-Hubbard model in the superfluid regime. In addition, the $3U$ resonance in the spectrum is a useful signature to characterize the density of the system. Therefore, the absorption spectroscopy is another experimental technique to study the density profile of the system besides the time-of-flight imaging. 

In conclusion, we have analysed in detail the dynamical response at zero temperature of the trapped ultracold bosons in an optical lattice subjected to lattice modulations. For the Mott-insulating system we identified several excitation processes. For the superfluid state, the presence of the harmonic trap induces slow-varying inhomogeneities, which can be understood within the LDA. All these unique properties can be used examine whether the Bose-Hubbard model is a good realization of the ultracold atom system in a parabolic trapping potential. On the other hand, if one believes that the model can explain all the physics in this system, absorption spectroscopy can be used as a powerful technique in revealing many basic features of the system, including its quantum state and density distribution.

\begin{acknowledgments}
We acknowledge part of financial support from HKSAR RGC Grant No. 701009. U.S. thanks the DFG for support through FOR 801. M.T. was supported by the Swiss National Science Foundation and by a grant from the Army Research Office through the DARPA OLE program. 
\end{acknowledgments}
\bibliography{mybib}

\begin{thebibliography}{29}
\bibitem [{foo()}]{footnote1}%
  \BibitemOpen
  \href@noop {} {}\bibinfo {note} {In all the cases below, a parabolic trapping
  potential with $V_t=0.003E_r$ is chosen.}\BibitemShut {Stop}%
\end{thebibliography}
\bibliographystyle{apsrev4-1}
\end{document}